\documentclass[aps,prl,twocolumn,superscriptaddress,showpacs]{revtex4}
\usepackage{graphicx}
\usepackage{epsfig}

\begin{document}

\newcommand{\new}[1]{{\em #1}\marginpar{\scriptsize new}}
\newcommand{\newB}[1]{{\em #1}\marginpar{\scriptsize newB}}
\newcommand{\newC}[1]{{\em #1}\marginpar{\scriptsize newC}}

\title{Triplet superconducting pairing and density-wave instabilities in
  organic conductors}        
 
\author{J. C. Nickel}
\affiliation{Regroupement Qu\'ebecois sur les Mat\'eriaux de Pointe,
  D\'epartement de physique, Universit\'e de Sherbrooke, Sherbrooke,
  Qu\'ebec, Canada, J1K-2R1 }
\affiliation{Laboratoire de Physique des Solides, CNRS UMR 8502,
  Universit\'e Paris-Sud, 91405 Orsay, France} 
\author{R.  Duprat}
\affiliation{Regroupement Qu\'ebecois sur les Mat\'eriaux de Pointe,
  D\'epartement de physique, Universit\'e de Sherbrooke, Sherbrooke,
  Qu\'ebec, Canada, J1K-2R1 }
\author{ C. Bourbonnais}
\affiliation{Regroupement Qu\'ebecois sur les Mat\'eriaux de Pointe,
  D\'epartement de physique, Universit\'e de Sherbrooke, Sherbrooke,
  Qu\'ebec, Canada, J1K-2R1 }
\author{N. Dupuis}
\affiliation{Laboratoire de Physique des Solides, CNRS UMR 8502,
  Universit\'e Paris-Sud, 91405 Orsay, France}
\affiliation{Department of Mathematics, Imperial College, 
180 Queen's Gate, London SW7 2AZ, UK}

\date{\today}
\begin{abstract} 
Using a renormalization group approach, we determine the phase diagram of an
extended quasi-one-dimensional electron gas model that includes interchain hopping,
nesting deviations and both intrachain and interchain repulsive
interactions. We find a close proximity of spin-density- and
charge-density-wave phases, singlet $d$-wave and triplet $f$-wave
superconducting phases. There is a striking correspondence between  our results  and recent puzzling experimental
findings in the Bechgaard salts, including the coexistence of
spin-density-wave and charge-density-wave 
phases and the possibility of a triplet pairing in the superconducting phase. 
\end{abstract}

\maketitle

Since the discovery of organic superconductivity made in the  Bechgaard
(TMTSF)$_2$X  salts more than two decades ago \cite{Jerome80}, the difficulty
of determining the origin of this phase remains one of the main focal points
of the   physics of  low dimensional  conductors.  The experimental weight
given recently to the hypothesis in favor of  a triplet rather  than singlet
superconducting phase  in these compounds \cite{Gorkov85,Lee97,Oh04,Lee02},
raises the problem of the microscopic conditions that can lead to triplet
pairing in correlated  quasi-one-dimensional (quasi-1D)  metals.  This
problem takes on particular importance in the Bechgaard salts series for which
superconductivity    in the phase diagram turns out to be surrounded by
dominant spin-density-wave (SDW) correlations as one moves along
the pressure, temperature or the magnetic field scale
\cite{Jerome82,Wzietek93,Chaikin96}. Repulsive intrachain interactions, that
are at the root of SDW correlations, are  well known to promote
unconventional singlet   pairing for superconductivity,  whenever nesting
properties of the quasi-1D  Fermi surface deteriorate under pressure
\cite{Emery86}. In the framework  of the quasi-1D electron
gas model with repulsive intrachain interactions, the application of the
renormalization group (RG) method, which allows to go  beyond mean-field and
RPA like  theories,  has shown  that for sufficiently large nesting
deviations, the interchain  electron pairing mediated by antiferromagnetic
fluctuations    becomes  invariably singular in the singlet interchain
`$d$-wave' channel \cite{Duprat01,Fuseya05}.   

On the other
hand, the extent to which  weaker  but yet present
charge fluctuations can act in expanding the range of pairing possibilities is
much less understood.   For repulsive intrachain  interactions,  it was found
from RPA like approaches that charge-density-wave (CDW) fluctuations 
enhance pairing
correlations in the triplet `$f$-wave' channel \cite{Kuroki01},  a
result that agrees with the Kohn-Luttinger mechanism for high -- odd --
angular momentum pairing induced by Friedel oscillations in  isotropic systems
\cite{Kohn65}.  Recent   RG calculations showed, however,  that for  repulsive
intrachain interactions in the quasi-1D case, the interchain `$f$-wave'  
correlations always
remain  subordinate to those of the `$d$-wave' channel \cite{Fuseya05}.  

Given that charges interact through the Coulomb interaction, not only
intrachain but also interchain interactions for electrons are present in
practice. The key role of interchain Coulomb  
interaction in the stabilization of a CDW ordered state in most 
Peierls quasi-1D
organic conductors has been made abundantly clear in the past
\cite{Gorkov74,Barisic85,Pouget89}. Their physical 
relevance in the Bechgaard salts has been borne out by the puzzling
observation of a CDW state that actually coexists with  SDW
\cite{Pouget96,Cao96}. In this Letter we give the 
first RG determination of the phase diagram for an extended quasi-1D electron
gas model that includes interchain hopping, nesting deviations  and both
intrachain and interchain repulsive interactions. The last interactions turn
out to  have  a 
sizable  impact on the structure of the phase diagram. Unexpectedly, we find
that  for a reasonably small amplitude of interchain interaction  the
`$d$-wave' superconducting (SC) ordered state is destabilized to the 
benefit of a
triplet `$f$-wave' phase with a similar range of $T_c$. The latter 
phase is preceded by
dominant antiferromagnetic correlations in the normal phase and by SDW order
at small  nesting deviations. In these conditions, the SDW state is 
found to be quite close in stability to a CDW phase.     

We consider weakly coupled conducting chains with a quasi-1D electron
dispersion $\epsilon(k_\parallel,k_\perp)-\mu = v_F(|k_\parallel|-k_F)
-2t_\perp\cos k_\perp-2t'_\perp\cos 2 k_\perp$, where $v_F$ is the
longitudinal Fermi velocity. The interchain hopping amplitude $t_\perp$
is small with respect to the longitudinal bandwidth $2\Lambda_0$, so
that the Fermi surface consists of two warped quasi-1D sheets around
$k_\parallel = \pm k_F$. The next-nearest neighbor hopping in the
transverse direction, $t_\perp'\ll t_\perp$, is used to parametrize deviations
from perfect nesting, which tend to suppress  the SDW
instability. We do not consider the
small interchain hopping in the third 
direction, which does not play an important role in our calculation, although
its existence is crucial for the stabilization of true long-range order at
finite temperature. Within the framework of an extended g-ology model, we
write the bare interaction amplitude as ($j=1,2,3$)
\begin{equation}
g_j(k'_{\perp1},k_{\perp2}',k_{\perp2},k_{\perp1}) = g_j + 2g^\perp_j
\cos(k'_{\perp 1}-k_{\perp 1}) ,
\end{equation}
where $k_{\perp1}\sigma,k_{\perp2}\sigma'$
($k'_{\perp1}\sigma,k_{\perp2}'\sigma'$)  are the transverse momenta and spins
of the two incoming (outgoing) particles.  
$g_1$ and $g_2$ correspond to backward and forward scattering, respectively,
and $g_3$ to longitudinal Umklapp processes with a lattice momentum 
transfer ${\bf  G}=(4k_F,0)$.  
The transverse momentum dependence comes from the nearest-neighbor interchain
interactions. Longer range (bare) interactions in the transverse direction are
expected to be very weak and are ignored. In this Letter, we
consider only the physically relevant case of repulsive
interactions ($g_j,g^\perp_j>0$). 
For the intrachain interaction constants, we take $\tilde g_1=0.32$, $\tilde
g_2=0.64$ and $\tilde g_3=0.02$, which falls into a realistic 
range of values compatible with various experiments  in the
Bechgaard salts \cite{Wzietek93,Emery86,Emery82,Schwartz98,Bourbon99}. 
The small (half-filling) Umklapp
process amplitude $\tilde g_3$ comes from the slight dimerization along the
organic chains \cite{Emery82}. $\tilde g_j=g_j/\pi
v_F$ and $\tilde g^\perp_j=g^\perp_j/\pi v_F$ are dimensionless interaction
constants. The bandwidth is taken to be $2\Lambda_0=30t_\perp$ with
$t_\perp=200$ K. Since the values of the interchain interaction amplitudes
$\tilde g^\perp_i$ are poorly known, we take them as free parameters with the
only constraint that they remain smaller than the intrachain interaction
amplitudes \cite{Saub76}. The latter condition is fulfilled in most
CDW systems \cite{Barisic85,Pouget89}. In order to minimize the number of
independent parameters, we restrict the discussion to the case $\tilde
g^\perp_1=\tilde g^\perp_2$ and $\tilde g^\perp_3/\tilde g_3=\tilde
g^\perp_1/\tilde g_1$; this turns out to be sufficient to understand the
global picture that emerges from our results. These  show no qualitative change over a sizable range of intrachain  interaction parameters. The key experimental control
parameters are temperature and pressure. Pressure affects 
$t_\perp$, $\tilde g_j,\tilde g^\perp_j$ and $t_\perp'$. However,  its main effect is to increase  $t'_\perp$ and therefore deteriorate
the nesting property of the Fermi surface.  

There are different ways to implement the RG approach to a quasi-1D system \cite{Duprat01,Honerkamp01b}. We
use the so-called one-particle irreducible (1PI) momentum-shell scheme as
developed in 
Ref.~\cite{Honerkamp01b}. One-loop RG equations for the two-particle vertices
and susceptibilities are solved numerically by dividing the Fermi surface into
$2\times 32$ patches. We retain only the $k_\perp$ dependence of the (running)
couplings $g_j(k'_{\perp1},k_{\perp2}',k_{\perp2},k_{\perp1})$. Various
instabilities of the normal phase are signaled by the divergence of the
corresponding susceptibilities.  

For $\tilde g^\perp_i=0$, the phase diagram has already been discussed in
Ref.~\cite{Duprat01}. When the nesting of the Fermi surface is nearly perfect
(small $t'_\perp$), the ground state is a SDW. Above a threshold value of
$t'_\perp$, the low-temperature SDW instability is suppressed and the ground
state becomes a $d_{x^2-y^2}$-wave superconducting (SC$d$) state  with an order parameter 
$\Delta_r(k_\perp)\propto \cos k_\perp$ [$r=+/-$ denotes the right/left sheet
of the quasi-1D Fermi surface]. 

\begin{figure} 
\epsfxsize 6cm
\epsffile[50 70 410 300]{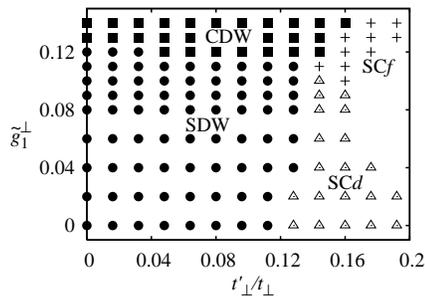}
\caption{$T=0$ phase diagram as a function of $t'_\perp/t_\perp$ and
  $\tilde g^\perp_1=\tilde g^\perp_2$ (with $\tilde g^\perp_3/\tilde
  g_3=\tilde g^\perp_1/\tilde g_1$).
Circles: SDW, squares: CDW, triangles: SC$d$ 
($\Delta_r(k_\perp)\propto\cos k_\perp$), crosses: SC$f$
  ($\Delta_r(k_\perp) \propto r \cos k_\perp$). }
\label{fig1}  
\end{figure} 

The $T=0$ phase diagram in the presence of interchain interactions ($\tilde
g^\perp_j>0$) is shown in Fig.~\ref{fig1}. 
For weak interchain interactions, we reproduce the phase diagram obtained in
Ref.~\cite{Duprat01}. As the interchain interactions increase, the region of
stability of the $d$-wave SC phase shrinks, and a triplet
$f$-wave (SC$f$) phase ($\Delta_r(k_\perp)\propto r\cos k_\perp$) appears next to the
$d$-wave phase for $\tilde g^\perp_1\simeq 0.1$. The
sequence of phase transitions as a function of $t_\perp'$ then becomes
SDW$\to$SC$d\to$SC$f$. For larger values of the interchain interactions, the
SC$d$ phase disappears and the region of stability of the $f$-wave
SC phase widens. In addition a CDW phase appears, thus giving the
sequence of phase transitions SDW$\to$CDW$\to$SC$f$ as a function of
$t'_\perp$. For $\tilde g^\perp_1 \gtrsim 0.12$, the SDW phase disappears.  
Note that for $\tilde g^\perp_1\simeq 0.11$, the region of
stability of the CDW phase is very narrow, and there is essentially a direct
transition between the SDW and SC$f$ phases.

\begin{figure}  
\epsfxsize 6cm
\epsffile[50 70 410 300]{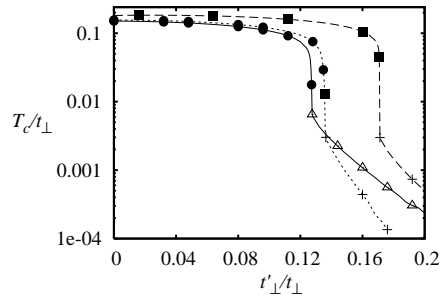}
\caption{Transition temperature as a function of $t'_\perp/t_\perp$ for
  $\tilde g^\perp_1=0$, 0.11 and 0.14, corresponding to solid, dotted, and dashed lines, respectively.} 
\label{fig2}
\end{figure} 

The transition temperature of the SDW phase is not very sensitive to the
values of the interchain interactions. The transition temperature of the SC
phase decreases for $\tilde g^\perp_1\lesssim 0.1$ (i.e. when the
SC  phase shrinks in the $T=0$ phase diagram) and increases for
$\tilde g^\perp_1\gtrsim 0.1$ (i.e. when the $T=0$ SC phase
widens). Our RG calculations yield $T_c\sim 30$ K for the SDW phase in the case of perfect nesting and
$T_c\sim 0.6-1.2$ K for the SC phase, in fair
agreement with experiments in the Bechgaard salts. 
Fig.~\ref{fig2} shows the  
transition temperature $T_c$ as a function of $t'_\perp$ for three different
values of the interchain interactions, $\tilde g^\perp_1=0$, 0.11 and 0.14,
corresponding to the three different sequences of phase transitions as a
function of $t_\perp'$: SDW$\to$SC$d$, SDW$\to$(CDW)$\to$SC$f$ and
CDW$\to$SC$f$.   

In the absence of interchain interactions, the effective interaction
  mediated by spin fluctuations is
attractive in the $d_{x^2-y^2}$- and $f$-wave channels.  It is  repulsive in the
$p_x$- ($\Delta_r(k_\perp) \propto r$) (at variance with a phenomenological  approach to superconductivity \cite{Lebed00}), the $p_y$- ($\sin k_y$) and $d_{xy}$-wave
  ($r\sin k_\perp$) 
channels. The $d$-wave correlations dominate over the $f$-wave ones as they
involve the three components of the spin fluctuations. 
The origin of the $f$-wave SC and CDW phases can be understood 
by considering the contribution of the $g^\perp_j$'s to the (bare) scattering
amplitudes in the singlet and triplet particle-particle channels, as well as in
the charge and spin channels. $g^\perp_1$ favors $(2k_F,\pi)$ CDW
and triplet SC 
fluctuations, but suppresses the singlet SC fluctuations; it does
not affect SDW fluctuations. There is also  an indirect effect, since CDW fluctuations, {\it via} the usual
mechanism of fluctuation exchange, enhance triplet SC
fluctuations and suppress singlet SC fluctuations.  A similar
analysis shows that $g^\perp_2$ has a detrimental effect on both singlet and
triplet nearest-neighbor chain SC pairing. Nevertheless, the RG
calculation shows that weak intrachain Umklapp processes (as present in the
Bechgaard salts) are sufficient to neutralize this effect  through an enhancement of both spin and charge fluctuations. As
for the interchain Umklapp processes ($g^\perp_3$), they oppose the effect
of $g_3$, thus pushing the occurrence of the CDW and SC$f$ phases to
slightly higher values of $\tilde g^\perp_1=\tilde g^\perp_2$.
   
\begin{figure}  
\epsfxsize 6cm
\epsffile[50 70 410 300]{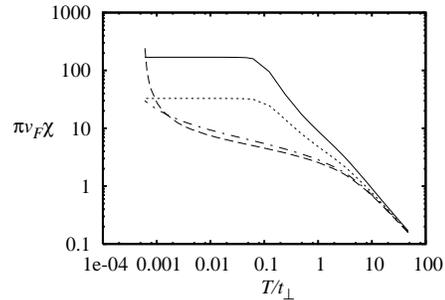}  
\caption{Temperature dependence of the susceptibilities in the normal phase
  above the SC$d$ phase [$t'_\perp=0.152 t_\perp$ and
  $\tilde g_1^\perp=0.08$]. The continuous
  line corresponds to SDW, the dotted line  
to CDW, the dashed line to SC$d$ and the dashed-dotted line to  
SC$f$ correlations which already show enhancement. } 
\label{fig3}
\end{figure}

\begin{figure}   
\epsfxsize 6cm
\epsffile[50 70 410 300]{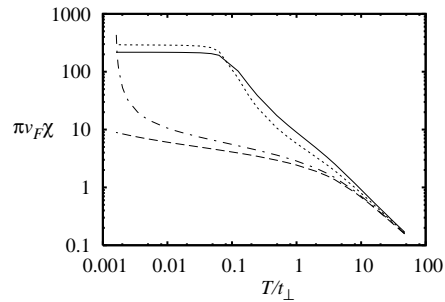} 
\caption{Temperature dependence of the susceptibilities in the normal phase
  above the SC$f$ phase [$t_\perp'=0.152 t_\perp$ and $\tilde
  g^\perp_1=0.12$]. }  
\label{fig4} 
\end{figure}

\begin{figure}   
\epsfxsize 6cm
\epsffile[50 70 410 300]{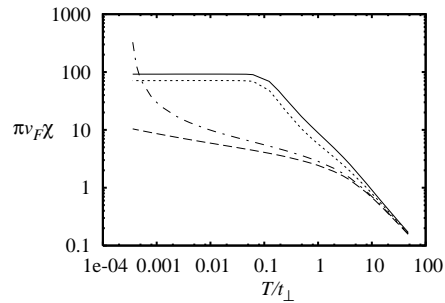} 
\caption{Same as Fig.~\ref{fig4}, but for $t'_\perp=0.176t_\perp$.} 
\label{fig5} 
\end{figure}

The RG approach also provides important information about the fluctuations in
the normal phase. It has already been pointed
out that the dominant fluctuations above the SC$d$ phase are SDW fluctuations
\cite{Duprat01}, as observed experimentally \cite{Wzietek93}. Although the
SDW fluctuations saturate below $T\sim t'_\perp$ where the 
SC$d$ fluctuations increase, the latter
dominate only in a very narrow temperature range above the SC
transition (Fig.~\ref{fig3}).  Above the SC$f$ and CDW phases, one expects strong CDW
fluctuations driven by $g^\perp_1$. Figs.~\ref{fig4}-\ref{fig5} show that for
$\tilde g^\perp_1 \sim 0.11-0.12$, strong SDW and CDW fluctuations coexist
above the SC$f$ 
phase. Remarkably, there are regions of the phase diagram where the
SDW fluctuations remain the dominant ones in the normal phase above the SC$f$
or CDW phase (Fig.~\ref{fig5}).

A central result of this Letter is the close proximity of SDW, CDW and SC$f$
phases in the phase diagram of a quasi-1D conductor
with {\it realistic} range of values for  the repulsive interactions. Although
this proximity is found only in a small range of
interchain interactions, there are several features of our results that
suggest that this part of the phase diagram is the relevant one for the
Bechgaard salts. i)
SDW fluctuations remain important in the normal phase throughout the whole
phase diagram;  they dominate  above the SC$d$ phase, and
remain strong (being sometimes even dominant) above the SC$f$ phase where
they coexist with strong CDW fluctuations, in accordance with observations
\cite{Wzietek93,Cao96}. ii) The SC$f$ and CDW phases stand 
nearby in the theoretical phase diagram, the CDW phase is always closely
following the SC$f$ phase when the interchain interactions increase. This
agrees with the experimental finding that both SDW and CDW coexist in the DW
phase of the Bechgaard salts \cite{Pouget96} and the existence,
besides SDW correlations, of CDW fluctuations in the normal state above the  
SC phase \cite{Cao96}. iii) Depending how one moves in practice 
in the phase diagram as a function of pressure in Fig.~1, our results are 
compatible with either a singlet $d$-wave or a triplet $f$-wave
SC phase in the Bechgaard salts. Moreover, we
cannot exclude that both SC$d$ and SC$f$ phases exist in these materials,
with the sequence SDW$\to$SC$d\to$SC$f$ under pressure. It is also possible 
that the SC$f$ phase,  not sensitive to the Pauli pair breaking effect, 
is stabilized by a magnetic field \cite{Shimahara00b,Fuseya05}. 
This would  provide an explanation for 
the existence of large upper critical fields
exceeding the Pauli limit \cite{Lee97,Oh04} and for 
the temperature independence of the NMR Knight shift in the
SC  phase \cite{Lee02}. Finally, the predicted existence of nodes in the SC gap for the $d$-  and  $f$-wave scenarios may appear in contradiction with the thermal conductivity and specific heat jump measurements for the  Bechgaard salt   (TMTSF)$_2$ClO$_4$,  which are apparently consistent with a nodeless order parameter \cite{note1}. Owing to the anion lattice superstructure of this compound, however, an `anion'  gap $\Delta_X \gg T_c$ must  be taken into account in the calculations so that a direct comparison with the RG method can be made.
\begin{acknowledgments}
J.C.N. is grateful to  the Gottlieb Daimler- und Karl Benz-Stiftung for partial
support.  C. B.  thanks  D. J\'erome, Y. Fuseya, M. Tsuchiizu, Y. Suzumura, L. G. Caron  and S. Brown  for  useful discussions.
\end{acknowledgments}


\end{document}